\documentstyle[12pt]{article}
\def\la{{\langle}}
\def\ra{{\rangle}}
\begin{document}

\begin{center}
{\Large \bf Phase space formalisms
of quantum mechanics
with singular kernel\vspace*{1.1cm}\\}
{\large R. Sala, J. P. Palao and J. G. Muga\vspace*{0.3cm}\\}
{\it Departamento de F\'{\i}sica Fundamental y Experimental\\
Universidad de La Laguna, La Laguna, Tenerife, Spain\vspace*{1.2cm}\\}
{\bf Abstract}
\end{center}
\vskip 3pt
\baselineskip 16pt
The equivalence of the Rivier-Margenau-Hill and  
Born-Jordan-Shankara phase space formalisms to the
conventional operator approach of quantum mechanics is demonstrated. 
It is shown that in spite of the presence of singular kernels
the mappings relating phase space functions
and operators back and forth are possible. 
$^{}$\vspace*{1.4cm}\\
Ref: Physics Letters A 231 (1997) 304, electronic version with permission
of Elsevier\\
e-mail address: JMUGA@ULL.ES
\vspace*{.2cm}  
\newpage 
\baselineskip 22pt
\pagestyle{plain}
%
%
\section{Introduction}
A wide class of quantum quasi-probability distributions $F$ of
position $q$ and momentum $p$ was studied by 
Cohen [\ref{Cohen}], the Wigner function [\ref{WW}] and the
Margeneau-Hill functions [\ref{MarHill}] being particular cases.
The choice among the $F$ functions representing the
quantum state of a system is similar to
the choice of a convenient set of coordinates. 
However in order to use safely one of these phase space functions 
it is necessary to show that the usual operator formulation of quantum
mechanics is equivalent to the corresponding phase space formalism.
In the later, in 
addition to the state, it is necessary to specify the functional 
form of the dynamical variables.    
In Cohen's approach each of the distributions $F$ is obtained   
with a particular kernel function $f$ from the quantum
density operator $\widehat{\varrho}$, $\widehat{\varrho}\to F$.      
(Operators are represented with an accent `` $\widehat{}$ ''
and the notation corresponds, as in [\ref{Cohen}],
to a particle moving in one spatial dimension.) 
Cohen gave as well the {\it quantization rule} $g\to\widehat{G}$ 
that associates with a
classical function $g(q,p)$ a quantum operator
$\widehat{G}(\widehat{q},\widehat{p})$  
in such a way that the expectation value of the operator 
can be written equivalently as a trace or as a phase space integral 
(All integrals in this work go from $-\infty$ to
$\infty$.) 
\begin{equation}\label{fa}
\la\widehat{G}(\widehat{q},\widehat{p})\ra=
{\rm tr} (\widehat{\varrho}\,\widehat{G})=
\int\!\!\int\, F(q,p)\, g(q,p)\, dq\,dp.
\end{equation}
The most frequently used quantization or ordering rules and
many others fit into this general scheme (in particular the 
rules by Weyl [\ref{Weyl},\ref{Moyal}],
Rivier [\ref{Rivier}], Born-Jordan [\ref{BJ}],
and the set of rules known as 
normal, antinormal, standard and antistandard
[\ref{Suda}-\ref{CG}]),
together with their associated phase space quasi-probability
distributions, see Table 1.   

In order to build a quantum formalism in phase space equivalent to 
the conventional operator approach, rather than
considering $\widehat{\varrho}$
and $g$ as primary, a different point of view is required 
where the primary objects are the state and observable
in operator form, $\widehat{\varrho}$ and $\widehat{G}$. Their
phase space representations, $F$ and $g$, are obtained from them
using $\widehat{\varrho}\to F$ and the inverse transformation of
the quantization rule, i.e., $\widehat{G}\to g$.    
Also in this case it is imposed that the expectation value of
$\widehat{G}(\widehat{q},\widehat{p})$ is given by 
(\ref{fa}), 
but now the function $g(q,p)$ is not necessarily equal to the 
classical magnitude; it is simply one the {\it images} or
{\it representations} of $\widehat{G}$ in phase space.

It is also possible to consider $F$ as a primary object. Then, the 
transformation $F\to \widehat{\varrho}$   
is required to obtain the corresponding density operator.
Even though it is not common practice to consider $F$ as primary 
in quantum mechanics, there are physical systems of practical importance
(semiconductor heterostructures) which are  
modelled in this fashion [\ref{Fren},\ref{MaHa}].
It is also of interest to note that classical statistical mechanics
is a theory formulated in terms of an
$F$ distribution (the classical distribution 
function) and properties $g$ (classical magnitudes), so that 
the transformations $F\to\widehat{\varrho}$ and $g\to\widehat{G}$
provide a set of equivalent operator formulations of classical 
statistical mechanics. Their potential applications are yet to be fully 
explored. One of them, using $f=1$ and associated
with the Weyl-Wigner formalism, was examined in [\ref{SM}].    
    
The (formal) full set of transformations that completes Cohen's two
original mappings,   
has been described by several authors
[\ref{MS}-\ref{Balazs}]. Whereas the transformations $g\to\widehat{G}$
and $\widehat{\varrho}\to F$ involve the kernel function $f$ in a
multiple integral, the inverse transformations involve $f^{-1}$. 
Agarwal and Wolf [{\ref{AW}] restricted their detailed study to 
mappings where the kernel function had no zeroes to avoid the
singularities of $f^{-1}$. In fact it has been generally believed 
that the inverse mappings cannot be performed in these cases
[\ref{Balazs}]. As a consequence the 
investigation or applications of some of the $f$ functions and
their associated quasi-probability distributions and
ordering rules have been scarce. Our main objective here is to
demonstrate that these singular kernels do not necessarily preclude
the existence of the inverse mappings and that indeed equivalent
phase space formalisms based on them can be constructed. 
In other words, we shall show that there is an
``inverse operator basis'', see (15) below,
in the form of operator valued distributions, 
associated with these kernels. The need to consider 
in general the mappings between operators and phase space from
a generalized function point of view
was already emphasized by Agarwal and Wolf [\ref{AW}].

The quasi-probability distributions in phase space are obtained from
$\widehat{\varrho}$ as 
\begin{equation}\label{510}
F(q,p)=\frac{1}{4\pi^{2}}
\int\!\!\int\!\!\int  
\bigg <u+\frac{\tau\hbar}{2}\bigg|\widehat{\varrho}
\bigg|u-\frac{\tau\hbar}{2}\bigg >
e^{-i[\theta(q-u)+\tau p]}f(\theta,\tau)\, d\theta\, d\tau\, du.
\end{equation}
Cohen noted that by imposing the condition 
\begin{equation}\label{511}
f(0,\tau)=f(\theta,0)=1, 
\end{equation}
the resulting $F$ function provides the correct ``marginal
distributions'' for $q$ and $p$.   
A group of $f$ functions 
and their corresponding quasi-probability distributions
are listed in Table 1 [\ref{other}]. 
The property (\ref{511}) is desirable
but this condition is not fulfilled by several useful quasi-probability
distributions [such as the $P$-distribution by Sudarshan [\ref{Suda}] and
Glauber [\ref{Gla}] or the $Q$ (or Husimi [\ref{Husi}])
distribution.]

The operator $\widehat{G}(\widehat{q},\widehat{p})$  
is given from the phase space function by   
\begin{equation}\label{517}
\widehat{G}(\widehat{q},\widehat{p})=
\frac{1}{4\pi^{2}}\int\!\!\int\!\!\int\!\!\int
g(q,p)f(\theta,\tau)e^{-i[\theta(q-\widehat{q})+
\tau (p-\widehat{p})]}\, 
dq\,dp\,d\theta\,d\tau.
\end{equation}
It is an exercise of Fourier transforms to obtain $g$ in terms of 
$\widehat{G}$ from (\ref{517}),    
\begin{equation}\label{520}
g(q,p)=\frac{\hbar}{2\pi}\int\!\! \int\!\! \int 
\bigg<u+\frac{\tau\hbar}{2}\bigg|\widehat{G}\bigg|u-
\frac{\tau\hbar}{2}\bigg>
\frac{e^{-i[\theta (q-u)+\tau p]}}{f(-\theta,-\tau)}
\, d\theta\, d\tau\, du
\end{equation}
Similarly 
\begin{equation}\label{Ftorho2}
\widehat{\varrho}=\frac{\hbar}{2\pi}
\int\!\!\int\!\!\int\!\!\int
F(q,p)f(-\theta,-\tau)^{-1}e^{-i[\theta(q-\widehat{q})+
\tau (p-\widehat{p})]}\, 
dq\,dp\,d\theta\,d\tau.
\end{equation}
The explicit expressions for all the transformations
in  equations (\ref{510}), (\ref{517}), (\ref{520})
and (\ref{Ftorho2}) can be summarized as 
\begin{eqnarray}
\label{Frho}
F&=&\Lambda[f]\,\widehat{\varrho}
\;\;\;\;\;\;\;\;\;\;
\;\;\;\;\;\;\;\;\widehat{\varrho}=\{\Lambda[f]\}^{-1}\,F
\\
g&=&h\Lambda[\tilde{f}^{-1}]\,\widehat{G} 
\;\;\;\;\;\;\;\;\;\;\;
\widehat{G}=h^{-1}\{\Lambda[\tilde{f}^{-1}]\}^{-1}\,g
\end{eqnarray}
where $\tilde{f}\equiv f(-\theta,-\tau)$.
          
Here we consider  
the transformation from $\widehat{\varrho}$ to $F$ as the reference
mapping   
represented by the ``operator'' (on the space of density operators)
$\Lambda[f]$ which depends functionally on $f$, see
(\ref{510}). The inverse operator $\{\Lambda[f]\}^{-1}$
acts on $F$ to provide the density operator $\widehat{\varrho}$. 
Note that $\widehat{G}\to g$ involves, except for a constant, the
same operation as $\widehat{\varrho}\to F$ but a different kernel, namely 
${\tilde{f}}^{-1}$ [\ref{Dahl}].    
      
These are of course formal results and for every $f$
it is necessary to study if these integrals exist and to determine 
domains where the transformations can be performed.  
Seemingly functions having zeroes, such as 
$f(\theta,\tau)=\cos({\theta\tau\hbar}/{2})$ or
$2\sin(\theta\tau\hbar/2)/\theta\tau\hbar$,
may be problematic because of the presence of the inverse of $f$ in
the integrands of the transformations $\widehat{G}\to g$ and
$F\to\widehat{\varrho}$.

It is also possible to relate 
operators and phase space using a framework complementary
to Cohen's [\ref{MS}-\ref{Balazs}]:  
Assume that there is an operator basis
$\widehat{B} (\widehat{q},\widehat{p};q,p)$ 
such that the operator $\widehat{G}(\widehat{q},\widehat{p})$ 
can be given as
\begin{equation}\label{521}
\widehat{G}=\int\!\!\int g_B(q,p)\widehat{B}(q,p)\, dq\, dp,
\end{equation}
where the ``coefficients'', $g_B(q,p)$, are, as before, the
{\it transform}, {\it image} or {\it representation}  
of the operator in that basis.    
For a basis $\widehat{B}$ to be practical it must have an inverse 
$\widehat{B}^-$ such
that [\ref{BB}]
\begin{equation}\label{522}
{\rm tr}[\widehat{B}^{-}(q,p)\widehat{B}(q',p')]
=\delta(q-q')\delta(p-p').
\end{equation}
If the density operator is expanded in the inverse basis,
\begin{equation}\label{rhoB}
\widehat{\varrho}=\int\!\!\int F_{B}(q,p)\widehat{B}^-(q,p)\, dq\, dp, 
\end{equation}
with expansion coefficients $F_{B}(q,p)$, 
using (\ref{522}) it follows that 
\begin{equation}
{\rm tr}(\widehat{G}\widehat{\varrho})=\int\!\!\int g_B F_{B}\,dq\,dp 
\end{equation}
and the coefficients (phase space representations of the state and
the observable) are obtained
from the operators by taking the traces
\begin{eqnarray}\label{FB}
F_{B}(q,p)&=&{\rm tr} (\widehat{\varrho}\widehat B)
\\
\label{gB}
g_{B}(q,p)&=&{\rm tr} (\widehat{G}\widehat{B}^-)
\end{eqnarray} 
In terms of $f$ these bases are formally given by  
\begin{eqnarray}\label{527}
\widehat{B}^-(q,p)&=&\frac{\hbar}{2\pi}\int\!\! \int 
e^{-i[\theta (q-\widehat{q})+\tau (p-\widehat{p})]}
\frac{1}{f(-\theta,-\tau)}\, d\theta\, d\tau
\\
\label{529}
\widehat{B}(q,p)&=&\frac{1}{4\pi^{2}}\int\!\!\int\
e^{-i[\theta(q-\widehat{q})+\tau(p-\widehat{p})]} 
f(\theta,\tau)d\theta\, d\tau.
\end{eqnarray}
From this perspective each of the phase space formalisms is based on
a certain ``coordinate system'' or ``basis'' that can be more or
less convenient depending on the case.  
But, as before, the presence of $f^{-1}$
in the expression for the inverse basis seems to impose a limitation
when $f$ has zeroes [\ref{Balazs}]. In this way one could be
prematurely tempted to discard, for example, phase space formalisms
based on the relatively common Rivier quantization rule
(or symmetrization rule) and the associated Margeneau-Hill function or
on the Born-Jordan quantization rule and the associated
Shankara distribution function [\ref{Shan}].
However the next section shows that the
zeroes of $f$ are not actually a problem.        
\section{Effect of zeroes of $f(\theta,\tau)$}
We shall give an
explicit expression 
of the transform of $\widehat{q}^n\widehat{p}^m$
for arbitrary nonnegative integer values of $n$ and $m$. 
According to equation (\ref{520}) the $f$-image of
$\widehat{q}^n\widehat{p}^m$ is obtained by the integral  
\begin{eqnarray}\label{ap1}
g(q,p)&=&g(q,p;\widehat{q}^{n}\widehat{p}^{m};f)
\equiv\frac{\hbar}{2\pi}\int\!\! \int\!\! \int 
\bigg<u-\frac{\tau\hbar}{2}\bigg|\widehat{q}^n\widehat{p}^m
\bigg|u+\frac{\tau\hbar}{2}\bigg>
\nonumber\\
&\times&
e^{i[\theta (q-u)+\tau p]}\frac{1}{f(\theta,\tau)}\,
d\theta\, d\tau\, du.
\end{eqnarray}
Introducing a closure relation in momentum this takes the form  
\begin{equation}\label{ap2}
g(q,p)=
\frac{1}{4\pi^2}\int\!\! \int\!\! \int\!\! \int  
\bigg(u-\frac{\tau\hbar}{2}\bigg)^{n}p'^{m}
e^{i\theta (q-u)}e^{-i\tau (p'-p)}\frac{1}{f(\theta,\tau)}\, 
d\theta\, d\tau\, du\, dp'.
\end{equation}
Making use of the integral expression of the Dirac delta and of its 
$m$-th order derivative $\delta^{(m)}$  the integral in $p'$ is solved,
\begin{equation}\label{ap3}
g(q,p)
=\frac{1}{2\pi(-i)^m}\int\!\! \int\!\! \int  
\bigg(u-\frac{\tau\hbar}{2}\bigg)^{n}
e^{i\theta (q-u)}e^{i\tau p}\frac{1}{f(\theta,\tau)}
\delta^{(m)}(\tau)\, d\theta\, d\tau\, du.
\end{equation}
The $\tau$-integral is carried out next. It is necessary to 
consider the  $m$-th derivative of the function 
\begin{equation}
\chi(\tau)=\bigg(u-\frac{\tau\hbar}{2}\bigg)^{n}
e^{i\tau p}\frac{1}{f(\theta,\tau)}. 
\end{equation}
By using  
Leibniz's formula for the $m$-th derivative of a product
twice and putting $\tau=0$ one obtains 
\begin{equation}
\frac{d^m\chi}{d\tau^m}=
\sum_{l=0}^{m}\left(\begin{array}{c}m\\l\end{array}\right)
\sum_{j=0}^l\left(\begin{array}{c}l\\j\end{array}\right)
(ip)^j\frac{n!}{(n+j-l)!}u^{n-l+j}(-\hbar/2)^{l-j}
\frac{d^{m-l}f^{-1}}{d\tau^{m-l}}\bigg|_{\tau=0},
\end{equation}
which is used to integrate over 
$\tau$ with the aid of the derivatives of the delta function.
Then the $u$-integral is carried out,
\begin{eqnarray}
g(q,p)&=&
(-1)^{n}
\sum_{l=0}^{m}\sum_{j=0}^{l}
\frac{1}{i^{m+n-l}}
\left(\begin{array}{c}m\\l\end{array}\right)
\left(\begin{array}{c}l\\j\end{array}\right)
p^j\frac{n!}{(n+j-l)!}(\hbar/2)^{l-j}
\nonumber\\
&\times&\int e^{i\theta q}
\frac{d^{m-l}f^{-1}}{d\tau^{m-l}}\bigg|_{\tau=0}
\delta^{(n-l+j)}(\theta)\,d\theta\,.
\end{eqnarray}
The derivatives of $f^{-1}$ with respect to $\tau$ at $\tau=0$
can be performed by using the formula for the derivative of arbitrary 
order, say $m-l$, of the inverse of a function
[\ref{Grand}]. This formula involves $f^{-1}(\tau=0)$, 
\begin{equation}
\frac{d^{m-l}f^{-1}}{d\tau^{m-l}}\bigg|_{\tau=0}=
(m-l)!(-1)^{m-l}\frac{1}{f^{m-l+1}(\tau=0)}D_{m-l}
\end{equation}
where
\begin{equation}
D_{m-l}\equiv\left|\begin{array}{cccccc}a_1&a_0&0&0&\cdots&0\\
a_2&a_1&a_0&0&\cdots&0\\
a_3&a_2&a_1&a_0&\cdots&0\\
\cdots&\cdots&\cdots&\cdots&\cdots&\cdots\\
\cdots&\cdots&\cdots&\cdots&\cdots&a_0\\
a_{m-l}&a_{m-l-1}&a_{m-l-2}&\cdots&a_2&a_1
\end{array}
\right|
\end{equation}
and
\begin{equation}
a_n=a_n(\theta)=\frac{1}{n!}\frac{d^{n}f}{d\tau^{n}}\bigg|_{\tau=0}
\end{equation}
Finally the integration over $\theta$ is performed using the 
derivatives of the delta function, and taking the condition 
(\ref{511}) into account. Again, Leibniz's theorem is used to 
arrive at the lengthy but explicit expression 
\begin{eqnarray}
g(q,p;\widehat{q}^{n}\widehat{p}^{m};f)&=&
\sum_{l=0}^{m}\sum_{j=0}^{l}\sum_{k=0}^{n-l+j}
(-1)^{m-j}\,(m-l)!\,i^{j-k-m}\,(\hbar/2)^{l-j}
\left(\begin{array}{c}m\\l\end{array}\right)
\left(\begin{array}{c}l\\j\end{array}\right)
\nonumber\\
&\times&
\left(\begin{array}{c}n-l+j\\k\end{array}\right)
\frac{n!}{(n+j-l)!}\,
p^j q^{n-l+j-k}\frac{d^k D_{m-l}}{d\theta^k}\bigg|_{\theta=0}
\label{Laecu}
\end{eqnarray}
where only mixed partial derivatives of the $f$ function at the 
origin ($\theta=0,\tau=0$) are to be considered.

We conclude that provided that Cohen's condition (\ref{511})
is satisfied and $f$ is 
analytical {\it at the origin} $g(q,p;\widehat{q}^n\widehat{p}^m;f)$
is well defined in spite of the possible zeroes of $f$.
The expression corresponding to the operators in
reverse order, $g(q,p;\widehat{p}^m\widehat{q}^n;f)$, is the same
except for a the change of sign $(\hbar/2)^{l-j}\to(-\hbar/2)^{l-j}$.   

Examples of images of
$\widehat{q}^2\widehat{p}^2$ for several $f$ using (\ref{Laecu})
are
\begin{eqnarray}
g(q,p;\widehat{q}^2\widehat{p}^2;f=1)&=&p^2q^2+2i\hbar
pq-\frac{\hbar^2}{2}
\nonumber\\
g\left[q,p;\widehat{q}^2\widehat{p}^2;f=
\cos(\theta\tau\hbar/2)\right]
&=&p^2q^2+2i\hbar pq
\nonumber\\
g\left[q,p;\widehat{q}^2\widehat{p}^2;f=2\sin(\theta\tau\hbar/2)/
(\theta\tau\hbar)\right]&=&p^2q^2+2i\hbar pq-\frac{\hbar^2}{3}
\nonumber\\
g\left[(q,p;\widehat{q}^2\widehat{p}^2;f=e^{-i(\theta\tau\hbar/2)}
\right]&=&p^2q^2
\nonumber\\
g\left[q,p;\widehat{q}^2\widehat{p}^2;f=e^{i(\theta\tau\hbar/2)}
\right] &=&p^2q^2+4i\hbar pq-2\hbar^2
\end{eqnarray}
It can be checked -via eq. (\ref{517}) with the
corresponding $f$ for each case- that
the operator obtained from these phase
space functions is indeed  $\widehat{q}^2\widehat{p}^2$, 
and that, once $f$ has been chosen, the relation between
operators and phase space images is biunivocal for arbitrary
(non-negative) values of $m$ and $n$.  
It is possible in summary to map into phase space functions
{\it at least} operators of $\widehat{p}$ and $\widehat{q}$ in
polynomial form or given by expansions in $\widehat{q}^n\widehat{p}^m$
or $\widehat{p}^m\widehat{q}^n$. A broad set of 
bounded operators admits such expansions [\ref{CG}].

Note that (\ref{Laecu}) is non-linear in $f$ so that 
if $f=af_1+bf_2$, in general $g_f\ne a g_{f_1}+b g_{f_2}$. 
In particular, the images obtained with 
the inverse of Rivier's rule are not given by the average of the
phase space representations using the inverses of the 
standard and antistandard rules. 

The arguments for demonstrating the feasibility of the inverse
transformation $F\to\widehat{\varrho}$ are analogous, based on expanding
$F$ in power series $\sum_{nm} b_{nm} q^np^m$ and transforming each
$q^np^m$ term independently to obtain the density operator as 
$\sum_{nm} b_{nm} \widehat{\varrho}_{nm}$.
Also in
this case the singularity is avoided due to the delta functions.
The result is       
\begin{eqnarray}
\nonumber
(\widehat{\varrho})_{nm}&=&
h\sum_{l=0}^m\sum_{k=0}^n\sum_{j=0}^{n-k}
\left(\begin{array}{c}m\\l\end{array}\right)
\left(\begin{array}{c}n\\k\end{array}\right)
\left(\begin{array}{c}n-k\\j\end{array}\right)
l!\,i^{l+k}(-1)^l 2^{k-n}
\\
&\times&\frac{d^k D_l}{d\theta^k}\bigg|_{\theta=0}
\widehat{q}^j\widehat{p}^{\,m-l}\widehat{q}^{\,n-k-j}\,.
\end{eqnarray}
Moreover, using (\ref{527}) and (\ref{529})
it is easy to check that (\ref{522}) is verified. In other words,
Rivier's rule, as well as other rules based on functions $f$ with
derivatives at the origin have an inverse basis in a generalized sense. 
Eq. (\ref{527}) is a symbolic expression that results from trying to fit 
the transformation (\ref{520}), whose calculation 
is possible as shown above, into the scheme of eq. (\ref{gB}). 
The fit requires a formal change in the order of integration. 
This generally illegitimate procedure is allowed when dealing with 
generalized functions. 
In this regard it is worth recalling that
the integral form of the a delta function [$(2\pi)^{-1}\int dx
\exp{(ixy)}$] is in fact a symbol that is not to be
interpreted ``literally''. The origin of this symbol is again a formal
change of integration when performing two successive Fourier
transformations. In the actual computation the order of integration 
is reversed, see the illustrative discussion in [\ref{BF}], or a
rigorous and more general analysis in
[\ref{Jones}] -especially sec. 7.9-.  
In the same vein, the actual order of integration when using 
(\ref{527}) is to be reversed so that it is never
performed over $\theta$ and $\tau$ {\it first}. This avoids the possible
singularities of $f^{-1}$ at the zeroes of $f$.

In summary, the number of phase space formalisms equivalent to
the conventional operator approach is broader than it had been generally
believed.
Zeroes of $f$ do not preclude the possibility of a set of
biunivocal transformations between operators and phase space
representations.         

{\bf Acknowledgments}

Support by Gobierno Aut\'onomo de Canarias (Spain) (Grant PI2/95)
and by Ministerio de Educaci\'on y Ciencia (Spain) (PB 93-0578)
is acknowledged.

\newpage

\centerline{\large{\bf References}}

\begin{enumerate}

\item\label{Cohen} L. Cohen, J. Math. Phys. 7 (1996) 781. 

\item\label{WW} E. Wigner, Phys. Rev. A 40 (1932) 749.

\item\label{MarHill} H. Margenau  and R. N. Hill, 
Progr. Theoret. Phys. (Kioto) 26 (1961) 722;
G. C. Summerfield and P. F. Zweifel, J. Math. Phys. 
10 (1969) 233;
R. I. Sutherland, J. Math. Phys.
23 (1982) 2389; S. Sonego, Phys. Rev. A
42 (1990) 3733.

\item\label{Weyl} H. Weyl, The Theory of Groups and Quantum
Mechanics, Dover, New York, 1950.

\item\label{Moyal} J. E. Moyal,  Proc. Cambridge Philos.
Soc.  45 (1949) 99.

\item\label{Rivier} D. C. Rivier, Phys. Rev.  83 (1957) 862.

\item\label{BJ} M. Born  and P. Jordan, Z. Phys.
34 (1925) 873. 

\item\label{Suda} E. C. G. Sudarshan, Phys. Rev. Lett
10 (1963) 277.

\item\label{Metha} C. L. Metha,  J. Math. Phys.
5 (1964) 677.

\item\label{CG} K. E. Cahill and R. J. Glauber, 
Physical Review 177 (1969) 1857.

\item\label{Fren} W. R. Frensley,  Phys. Rev. B 36
(1987) 1570; Rev. Mod. Phys. 62 (1990) 745. 

\item\label{MaHa} R. K. Mains  and G. I. Haddad, 
J. Comp. Phys. 112 (1994) 149, and references therein  

\item\label{SM} R. Sala  and J. G. Muga,  
Phys. Lett. A  192 (1994) 180.

\item\label{MS} S. P. Misra and T. S. Shankara,  J. Math.
Phys. 9 (1968) 299. 

\item\label{AW} G. S. Agarwal and E. Wolf,  Phys. Rev. D
2 (1970) 2161; 2187; 2206.  

\item\label{SW} M. D. Srinivas and E. Wolf, Phys. Rev. D
11 (1975) 1477. 

\item\label{Balazs} N. L. Balazs and B. K. Jennings,
Phys. Rep. 104 (1984) 347.

\item\label{other} Other choices of $f$ 
may be found in [\ref{Cohen},\ref{CG},\ref{Balazs}],
and in  
L. Cohen and Y. I. Zaparovany,  J. Math.
Phys. 21 (1980) 794; L. Cohen,  Ann. New York Acad. Sci.
480 (1986) 283; A. K. Rajagopal and S. Teitler, Pramana J. Phys.
33 (1989) 347.

\item\label{Gla} R. J. Glauber, Physical Review
131 (1963) 2766.

\item\label{Husi} D. Lalovic, D. M. Davidovic and N. Bijedic, 
Phys. Rev. A  46 (1992) 1206.

\item\label{Dahl} The kernels $f(\theta,\tau)$ and $f(-\theta,-\tau)^{-1}$
define a set of dual phase space representations,
see J. P. Dahl, in: H. D. Doebner,
W. Scherer and F. Schroeck, Jr. (Eds.), 
Classical and Quantum Systems, Foundations
and Symmetries, World Scientific, Singapore, 1993, pg 420. 

\item\label{BB} Do not miss $\widehat{B}^-$ with the 
reciprocal basis of $\widehat{B}$, defined by 
$
\widehat{B}^{-1}\widehat{B}=\widehat{B}\widehat{B}^{-1}=\widehat{1}.
$

\item\label{Shan} T. S. Shankara, Progr. Theoret. Phys.
(Kyoto) 37 (1967) 1335.

\item\label{Grand} I. S. Grandshteyn  and M. Ryzhik, 
Table of Integrals, Series, and Products,
Academic Press, San Diego, 1994.

\item\label{BF} F. W. Byron and R. W. Fuller, Mathematics of 
Classical and Quantum Physics, Addison, Reading MA, 1969, pg 248.

\item\label{Jones} D. S. Jones, Generalised functions,
McGraw-Hill, London, 1967.

\end{enumerate}
\newpage
\noindent TABLE 1\\
%

\begin{flushleft}
\begin{tabular}{|c|c|c|}\cline{1-3}
f&F&quantization rule
\\\hline\hline
1&Wigner&Weyl
\\\cline{1-3}
$\cos(\theta\tau\hbar/2)$&Margenau-Hill&Rivier(symmetrization)
\\\cline{1-3}
$2\sin(\theta\tau\hbar/2)/\theta\tau\hbar$&Shankara&Born-Jordan
\\\cline{1-3}
$e^{-i(\theta\tau\hbar/2)}$&Kirkwood $f_K^+$&standard
\\\cline{1-3}
$e^{i(\theta\tau\hbar/2)}$&Kirkwood $f_K^-$&antistandard
\\\cline{1-3}
$e^{\frac{\hbar}{4}[(\tau\lambda)^2+(\theta/\lambda)^2]}$
&P-function(Sudarshan-Glauber)&normal
\\\cline{1-3}
$e^{-\frac{\hbar}{4}[(\tau\lambda)^2+(\theta/\lambda)^2]}$
&Q-function(Husimi)&antinormal
\\\cline{1-3}
\end{tabular}
\vspace*{.5cm}\\
TABLE 1  CAPTION: 
$\lambda$ is a real parameter different
from zero.

\end{flushleft}

\end{document}